# BIBLIOMETRIC ANALYSIS OF AGILE SOFTWARE DEVELOPMENT

F. Almeida[1*]

[1]Faculty of Engineering of Oporto University, INESC TEC, Porto, Portugal

*ABSTRACT*

*Agile methodologies are currently considered one of the main paradigms of software development. Its study, from a scientific point of view, has deserved prominence in recent years by the scientific community related to the area of software engineering. This study intends to perform a bibliometric analysis of the quantity, characteristics and scope of the most relevant studies published in this area of knowledge. The findings indicate that the number of studies published from 2010 to 2016 significantly increased, having reached a peak in 2015. The study identifies the main journals and conferences in the field and we also concluded that the majority of published studies are literature reviews of agile software development, and qualitative and quantitative research methods have identical number of publications.*

## 1. INTRODUCTION

Agile methodologies are an alternative to traditional project management practices. They were born in the context of software development, but today they can be applied to any type of project in different fields of activity. Agile methodologies have been helping many teams deal with unpredictability within a project through incremental deliveries and iterative cycles.

Agile approaches seek to promote a project management process that encourages frequent inspection and adaptation. It is a philosophy that ultimately encourages greater teamwork, self-organization, frequent communication, customer focus and value delivery. Basically, agile approaches are a set of effective practices that are designed to enable fast and high-quality product delivery. In this sense, agile offers a business approach that aligns project development with customer needs and business goals.

According to its promoters, the agile manifesto has twelve principles and four values that intend to give significance for the following aspects:
- Individuals and interactions over processes and tools;
- Working software over comprehensive documentation;
- Customer collaboration over contract negotiation;
- Responding to change over following a plan.

The agile manifesto does not reject processes and tools, documentation, contract negotiation or planning, but simply shows that they have a secondary importance when compared to individual and interactions, software running, client collaboration, and quick responses to changes.

---

[*] Corresponding Email: almd@fe.up.pt

There are already in the literature some studies that analyze the bibliometric impact of agile software development. However, none of them analyze together the evolution of the number of articles published in journals and conferences with their adopted research methodology. In addition, some of these studies focus exclusively on one of the most used methods in an agile environment such as Scrum or focus exclusively on one of the components of the agile methodology, such as risk management. Thus, this paper intends to systematically analyze the quantity, characteristics and scope of the published studies indexed by Web of Science and Scopus during the period 2010-2016. The paper is organized as follows: First, we perform a review of the most relevant related works on the field of agile software development. Then, we present the adopted methodology to perform a bibliometric analysis, followed by the presentation and discussion of the main findings. Finally, we draw the conclusions of our work.

## 2. RELATED WORKS

The field of software engineering has experienced a very significant growth and a strong interest from the scientific community. One of the most relevant and complete studies in this field was carried out by Garousi & Mäntylä (2016), which characterizes the most predominant authors, research topics and active countries in software engineering. The study identified that around 6000-7000 papers are published every year, being the United States, China and United Kingdom the countries that contribute most to this statistic. The most relevant research topics in software engineering identified by the authors are: (i) web services; (ii) mobile and cloud computing; (iii) industrial/case studies; (iv) source code; and (v) test generation.

Most studies tend to focus their analysis on a sub-domain of software engineering. In this area, four studies that adopt a bibliometric approach deserve particular attention. Adams & Pinto (2005) conducted a bibliometric analysis about the importance given to the topic of risk management in software engineering. The results indicated that until that date the number of scientific studies addressing risk management is very low and, therefore, there is a need for additional focused research in software risk management. A similar study in the same field was conducted by Tavares et al. (2017), after twelve years since the realization of the first study, but the obtained results are similar and confirmed that there are still few scientific studies about risk management in Scrum projects. Karg et al. (2011) perform a systematic review on the field of software quality costs, reaching the conclusion that the majority of articles in the field does not empirically validate their findings because they typically do not target multiple companies. Anyway, it was possible to conclude that prevention costs have gained the least attention, in spite of their big cost impact. Freitas & Souza (2011) performed also a bibliometric analysis considered four dimensions (i.e., publication, sources, authorship, and collaboration) on search based software engineering. The study covered 740 publications in the field and it was possible to identify a significant growth (i.e., around 20%) of this research domain between 2001 to 2010.

Lean practices, whose essence is based on the ability to eliminate waste and solve problems in a systematic way with strong emphasis on creating value for the customer, have gained wide acceptance in the market particularly in the area of manufacturing. However, these lean practices progressively migrated to other areas, such as the services. At this level, we can also assist to significant improvements in terms of service quality while reducing the costs (Vignesh et al., 2016). In the software field, we can also find several benefits, such as reduced lead time, higher productivity, lower cost and higher customer satisfaction. Rodríguez et al. (2014) also carried out a case study in this field by looking how lean can be combined with agile methods to enhance software development processes. The findings evidenced numerous compatibilities between lean and agile, and it was possible to realize that lean thinking has brought new elements to software development, such as Kanban, work-in-progress limits, a "pull" and "less waste" oriented culture, and a more collaborative and transparency development. On the other side, the need for management commitments, requirements of change in the organization and the office space and incompatibility with traditional project management was identified as main impeding factors for introduction of lean software development (Jonsson, 2012).

Gustavsson (2016) carried out an analysis, adopting an inverse perspective, in which he tries to demonstrate that the agile principles of software development can be used and adapted to other areas of activity. The results of this study confirm benefits in terms of teamwork, customer interaction, productivity and flexibility in different areas, such as library management or strategic management. Additionally, Christie et al. (2012) performed a literature review on prototyping strategies. The study grouped prototyping approaches into several categories with similar characteristics and found that prototyping strategies are used in a wide variety of different areas, such as business or engineering perspective.

Agile software development methodologies are currently one of the most relevant and emerging issues in the field of software engineering. Several studies that systematize the main lines of research in this field have been carried out. Dyba & Dingsøyr (2008) identified four themes: (i) introduction and adoption, (ii) human and social factors, (iii) perceptions on agile methods, and (iv) comparative studies. Furthermore, they look at the benefits and limitations of agile software methods. Dingsøyr et al. (2012) provided an overview of the last decade (2002-2012) of agile software development. It was possible to conclude that agile software development is a very stimulating research area in terms of publications in scientific journals and international conferences. Additionally, the study revealed that the three most important countries in terms of publications are: (i) USA; (ii) Canada; and (iii) Germany.

"Agile" is seen today in the software industry as more than just a methodology. In reality, it is a philosophy of work that implies significant changes in the management of teams and people. Sheuly (2013) presented a systematic review of literature on agile project management (APM). The findings showed that the majority of studies in this field looks at how practitioners implemented agile methods. The study also revealed that there are several APM methods, such as Scrum, extreme programming (XP), Crystal clear method, DSDM and lean programming. More recently Mishra et al. (2015) performed a comparative analysis of agile software development methodologies, considering the most popular agile methodologies, such as Scrum, XP, and Kanban. Lechler & Yang (2017) also explored the role of project management in agile software development. They concluded that the agile academic discussion has been mainly unidirectional, which typically examines the challenges created by agile methodologies in project management. On the contrary, there are deficiencies in the analysis of how different project management techniques can be applied in agile methodologies.

Hossain et al. (2009) focused their analysis on the adoption of Scrum in Global Software Development (GSD) projects. They identified 20 primary research papers in this field and they concluded that Scrum practices are not totally suitable to support globally distributed software development teams, where typically appear significant issues in terms of communication, coordination and collaboration processes. Rizvi et al. (2015) conducted a study to understand the reasons and conditions that lead to the adoption of distributed agile software engineering practices. The findings found difficulties faced by teams in terms of zone difference, knowledge of resources and lack of infrastructure. In terms of proposals that facilitate work in geographically distributed teams they found recommendations of having a good infrastructure in place for communication, encouraging team members to engage in formal and informal communications, and good tools that could enhance the collaboration between team members.

The adoption of agile methodologies in certain domains and sectors of activity has also been the subject of research. Vacari & Prikladnicki (2015) looked to the adoption of agile methods in the public sector. They realized that the number of structured studies about it is very reduced. However, despite that, they concluded that agile projects can be adopted in the public sector and, for that, it is suggested to start the adoption of agile with the support of senior leaders on important pilot-projects. Some barriers were identified, such as the existence of a plan-drive methods, big deliveries and lack of experience using agile methods. Hajou et al. (2014) looks how the pharmaceutical industry and agile software development methods conflict. This research concluded that the application of agile methods in the pharmaceutical is almost nonexistent, which typically causes most projects to exceed budget.

Chow & Cao (2008) performed a survey to identify the critical success factors in agile software projects. This study identified three critical success factors for agile software development projects: (i) delivery strategy; (ii) agile software engineering techniques; and (iii) team capability. Wan & Wang (2010) also performed a similar study and identified the following three critical success factors: (i) importance of education and training; (ii) promotion of an agile culture of mutual trust and cooperation; and (iii) attention to the design phase of the application. El Hameed et al. (2016) classified the critical success factors of agile software development using a mind map. This study proposed the following six groups of critical factors: (i) process; (ii) organization; (iii) technical; (iv) people; (v) project; (vi) product. All the elements identified by the other studies appear within each of the identified groups.

Silva et al. (2011) focus their study in the integration of agile software development with user-centered design (UCD) approaches. The findings of this study provide two relevant conclusions: (i) the focus on integrating agile methods and UCD should be on design as well as on usability evaluation; and (ii) none of the identified studies validates the integration of UCD and Agile with controlled experiments. Bano & Zowghi (2013) performed a systematic literature review about the contribution of user involvement in the software development process. The work reviewed 87 studies and found that around 68% of them considered that user involvement positively contributes to system success.

The process of gathering and identification of requirements using agile development methodologies is another pertinent theme. Inayat et al. (2015) identified 21 papers that discuss agile requirements engineering. The study identified 17 practices of agile requirements engineering and some challenges that have been solved, but also introduced, by the practice of agile requirements engineering. On the other hand, Schön et al. (2017) identified 27 relevant papers that address agile requirements engineering and look in detail for the role of stakeholder and user involvement, data gathering, documentation and identification of non-functional requirements.

Other areas of the software development cycle have also been studied by several authors. For example, in the maintenance and support area, the study conducted by Naz & Khan (2015) performed a literature review of rapid application development techniques. The authors found that product maintenance and support phase of the Scrum have been little explored areas, hence they propose a novel model that focuses on the maintenance phase of Scrum. Tarwani & Chung (2016) performed a review on agile software maintenance techniques and tools. The authors concluded that the traditional maintenance phase of the software lifecycle has been addressed by agile methodology in terms of visibility reduced cost.

### 3. METHODOLOGY

The aim of this study is to analyze the evolution and the characteristics of the publications within agile software development methodologies. For that, we considered all articles published from 2010 that are indexed in INSPEC. Instead of using the Web of Science or Scopus that focus is targeted in only top journals and conferences, the use of INSPEC let us to analyze a more diverse range of publications. Currently, INSPEC has over 16 million abstracts in the field of physics and engineering. This approach establishes the state of current knowledge in the field and allows us to perceive the different dimensions in a given area of knowledge. We adopted the methodology proposed by Royle et al. (2013), which let us establish a process and define guidelines for the accomplishment of bibliometric studies.

As a starting point we used the study conducted by Dingsøyr et al. (2012) that analyzes the evolution of the number of publications on agile software development from 2001 to 2010. That study considers only two dimensions: (i) number of journal articles; and (ii) number of conference papers. In our study, we added a new dimension for analysis entitled the number of books. Then we used the INSPEC database to search for articles and books published between 2010-2016. To identify the largest number of articles in the field, we considered three search terms: (i) agile software development; (ii) scrum; (iii) extreme programming; (iv) test-driven development; and (v) Kanban software. We eliminated duplicated entries to avoid duplicating counting. These keywords were used in considering the popularity of agile methods and practices (Partogi, 2016).

As shown in Figure 1, the number of published articles in the field increased from 2010 to 2016 by approximately 79,4%. However, there were some oscillations in this period, particularly in the last considered year. It should also be mentioned that the total number of publications has suffered a slight decrease, in order of 11,2%, from 2015 to 2016. It was precisely in the year of 2015 that there was a peak of scientific publications in the field of agile methodologies. The contribution of publications in journals has been constantly the most relevant dimension among the years, followed by articles published in conference proceedings and books. Lastly, it should be highlighted that the publication of books and book chapters has gained a consistent and impressive increase over the past six years (around 223%).

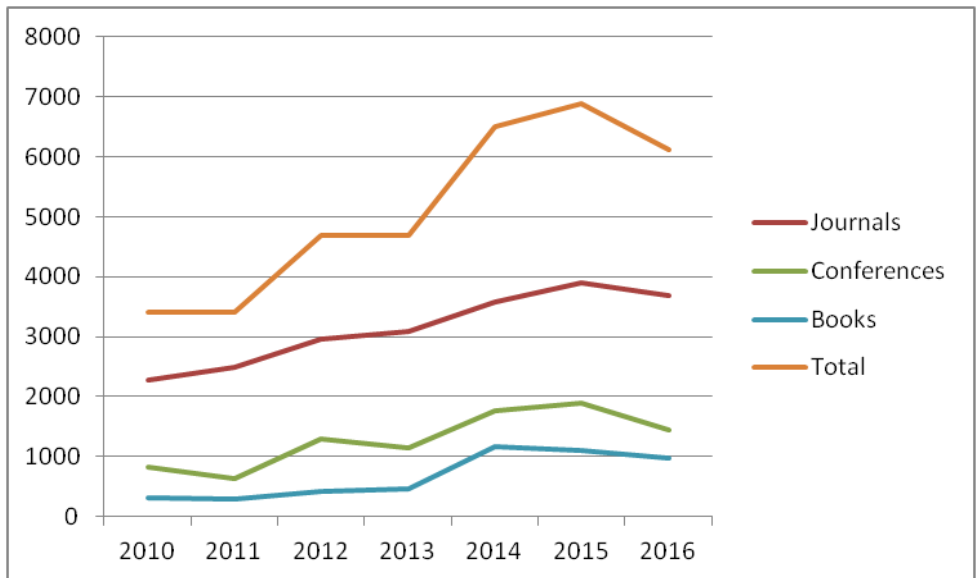

Figure 1. Publications on agile software development from 2010 to 2016

Our objective was to establish what is known about: (i) the most relevant journals and conferences in the field; (ii) the most common subjects associated with agile software development; and (iii) the adopted methodology.

As a final step in our analysis, we systematically compared the results obtained from studies in the field of agile software development with the findings obtained by other authors in the field. We critically analyze the results in order to find answers and to draw guidelines for further studies in the field.

## 4. FINDINGS AND DISCUSSION

The findings are organized in the following three dimensions:
1. Analysis by journals and conferences - identification of the main journals and conferences where the scientific studies in the field of agile methodologies are published;
2. Analysis by subject - seek for the keywords associated with scientific publications on agile methodologies;
3. Analysis by type of research method - identification of the type of scientific methodology adopted in each study.

### 4.1 Analysis by journals and conferences

Figure 2 synthesizes the number of journal publications from 2010 to 2016. The analysis includes only the top 10 journals and reveals that IEEE Software has the largest number of papers, followed by the International Journal of Production Research, and Information and Software Technology. It is also important to mention the acceptance of several articles in the area of software development using agile methodologies in several scientific journals, which their main focus are not in the field of software engineering. For instance, journals in the field of general science, industrial engineering, human-machine interaction, and educational sciences have also devoted some of their volumes to the analysis of this topic. This happens because agile methodologies are a multidisciplinary theme with several areas of application, namely in the management and industrial fields.

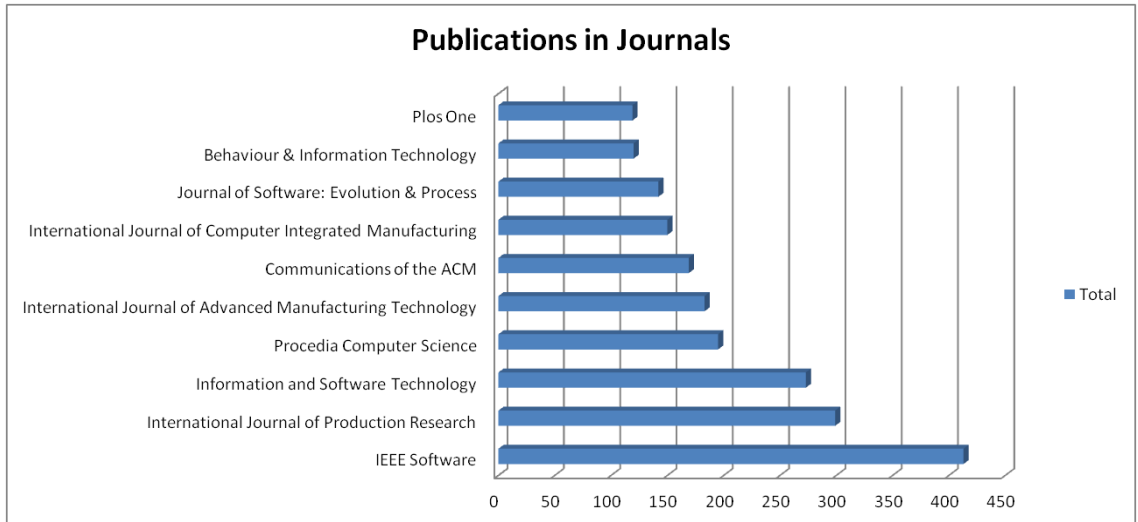

Figure 2. Number of papers in scientific journals

Figure 3 illustrates the number of scientific conference publications on the same period of time. Like in the previous scenario, only the top 10 scientific conferences were considered. The various editions of the same conference were aggregated in just one entry to allow a full counting of the total number of publications. It is the case of the Agile Conference and Hawaii International Conference. The analysis reveals that the three most relevant conferences in the field are the ACM International Conference, International Conference on Software Engineering (ICSE), and the Agile Conference.

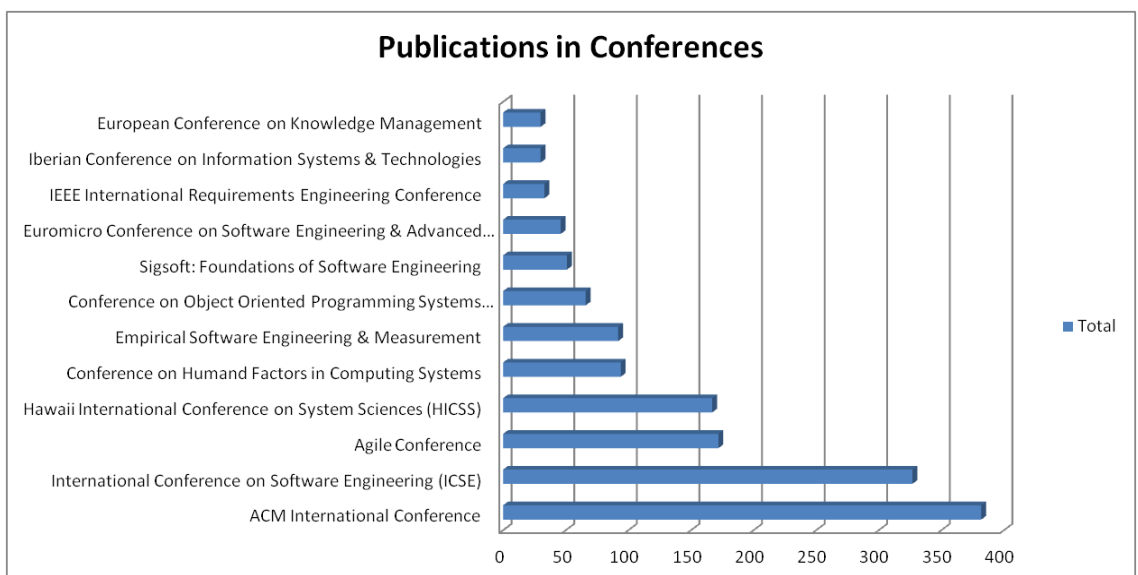

Figure 3. Number of papers in scientific conferences

**4.2    Analysis by subject**

An analysis of the most common subjects associated with scientific publications in the field of agile software engineering was also performed. The five keywords used in the process of searching for publications and all the combinations of the words "software" and "development" were eliminated. Looking at Figure 4 it is possible to realize that not surprisingly the three most relevant associated subjects are related to computing, information technology, and project management. However, other popular subjects appear to have also significant importance, such as decision making, technological innovations, supply chain management and knowledge management. It is important to highlight the importance of agile methodologies in several areas of engineering and social sciences. At this level, there are several studies where agile methodologies are used in the field of process improvement, human resources management and in the implementation of innovation policies.

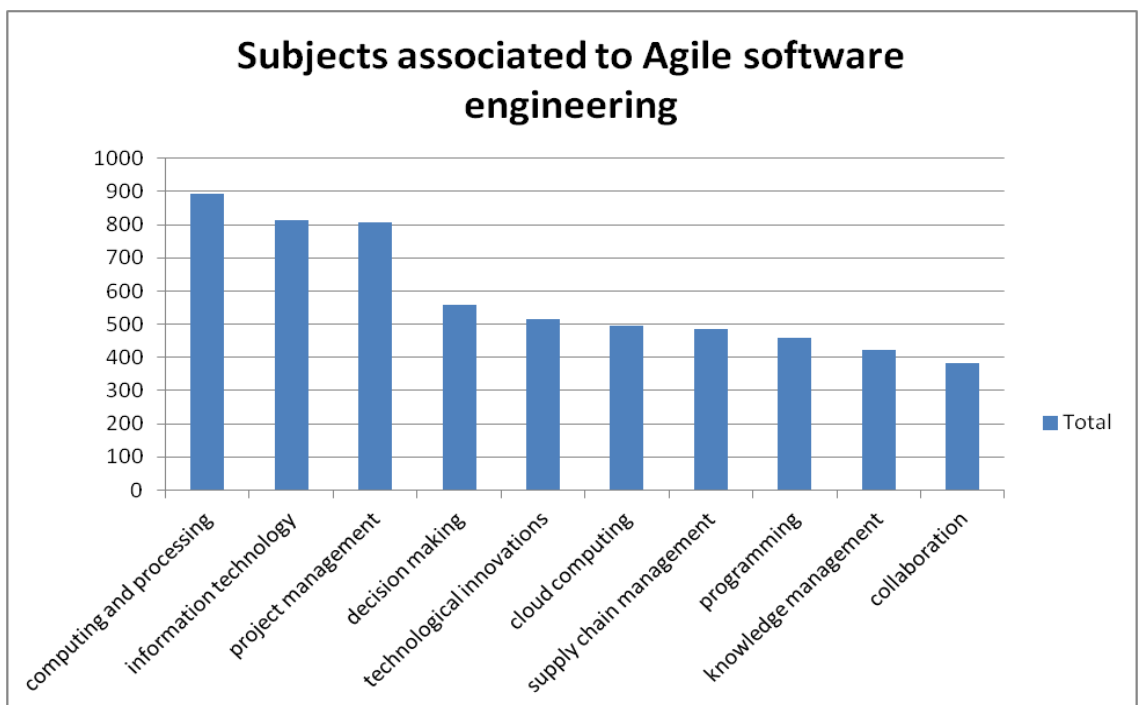

Figure 4. Number of publications by each associated subject

**4.3    Analysis by type of research method**

Scientific research studies can be organized into four broad types: (i) literature review; (ii) qualitative research; (iii) quantitative research; and (iv) mixed methods research. The literature review is applied in areas where there are already a significant number of publications, in order to produce a synthesis of the primary studies in the field, evidencing their conclusions and discussing the differences between these primary studies. Qualitative studies allow the researcher to explore a given area of knowledge. Therefore, the researcher describes, interprets, and explains a phenomenon of interest. On the other hand, quantitative studies aim to evaluate objectively a given area of knowledge and seek measurable relationships among variables to test and verify their study hypotheses. Finally, mixed methods research combines both qualitative research and quantitative research in the same study (Creswell, 2017; Swanson & Holton, 2005; Warfield, 2010).

In order to perform a systematic search for each type of research method described previously, we considered the keywords presented in Table 1 as search terms. For its part, the results of this research are shown in table 2. For each research method, we considered the number of publications (No.) and its corresponding percentage (PCT.) in journals and conferences. A total of 29667 publications were mapped into the four research method categories. When comparing this number with the total number of publications from 2010 to 2016, we verified that around 6000 studies were not possible to be cataloged. This happens because the methodology adopted in these articles was not included in the keywords of the document or the adopted methodology doesn't fit within the search terms defined in Table 1. Looking at Table 2, we can verify that there are a greater number of publications using a literature review approach both in journals and conferences. The number of published studies adopting a qualitative or qualitative is similar. However, the distribution of research methods for the type of publication is not totally uniform. For instance, we have around 82,37% of the mixed methods studies published in journals, while it is lower for the other types of publications (e.g., for literature review is 79,99%; qualitative research studies represent 78,26%, and it is 79,15% for quantitative research studies.

Table 1. Search terms used to map each research method

| Research method | Search terms |
|---|---|
| Literature review | literature review, state of the art |
| Qualitative research | qualitative, case studies, focus groups, ethnography, grounded theory |
| Quantitative research | quantitative, survey, correlational, experimental research |
| Mixed methods research | mixed methods, sequential explanatory, sequential exploratory, concurrent nested |

Table 2. Number and percentage of studies by each research method

| Research method | Journals | | Conferences | | Total |
|---|---|---|---|---|---|
| | No. | PCT. | No. | PCT. | No. |
| Literature review | 9672 | 32,60% | 2420 | 8,16% | 12092 |
| Qualitative research | 5322 | 17,94% | 1478 | 4,98% | 6800 |
| Quantitative research | 5349 | 18,03% | 1409 | 4,75% | 6758 |
| Mixed methods | 3309 | 11,15% | 708 | 2,39% | 4017 |

Finally, Figure 5 shows the total number of publications distributed by each research method. Literature review studies represent around 41% of the total number of publications, while mixed methods represent only 13%. The total number of studies that perform a qualitative research or adopt a quantitative research method are similar (23% for both situations).

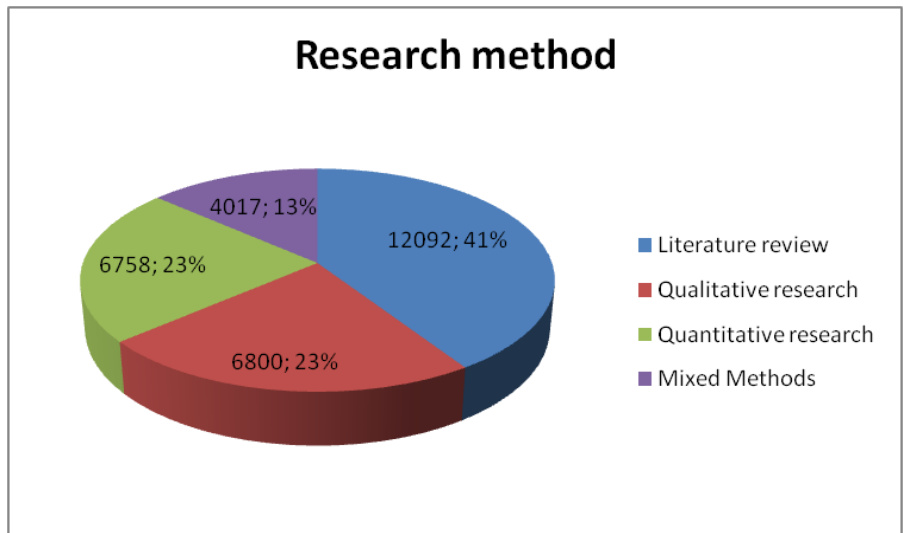

Figure 5. Total number of publications by each research method

## 5. CONCLUSION

Agile is a new way of developing and managing software processes that follow an interactive and incremental execution approach to complex, uncertain and unpredictable projects. The agile software development process arose due to increased marketing pressures for innovation, increased productivity and flexibility, and improved software quality. In this sense, the agile software methodology allows us to improve the way we are developing software with a principal focus on customer satisfaction.

It has not been just the business community to focus on agile methodologies. Likewise, there is also a significant interest from the scientific community on this model of software development. From 2010 to 2016, we assisted to a significant growth in the total number of publications, having reached a peak in the year of 2015. The publication of books and book chapter on agile methodologies has been one of the areas that have grown more in recent years. We also identified that the scientific journals with the largest number of publications in this field are: (i) IEEE Software; (ii) International Journal of Production Research; and (iii) Information and Software Technology. On the other hand, the conferences with the largest number of accepted papers in this field are: (i) ACM International Conference; (ii) ICSE; and (iii) Agile Conference. Finally, it was possible to conclude that the majority of published studies are literature reviews around the various dimensions of agile methodologies, which are more predominant in journals. On the other side, mixed methods research is still an emergent methodology representing around only 13% of the published studies.

When comparing the findings obtained from our study in relation to the study conducted by Dingsøyr et al. (2012), we got significant changes in the total number of publications. The use of the INSPEC database allowed us to map a much higher number of scientific papers and journals became the largest source of data. However, the journals and conferences with the largest number of publications in the field do not undergo significant changes. Finally, it was not possible to compare the number of subjects associated with agile software development and the research methods adopted by them, because these elements were not previously analyzed by Dingsøyr et al. (2012).